# ANALYSIS OF DIFFERENT PRIVACY PRESERVING CLOUD STORAGE FRAMEWORKS

Rajeev Bedi[1] Mohit Marwaha[2] Tajinder Singh[3] Harwinder Singh[4] Amritpal Singh[5]

[1]Department of Computer Science and Engineering, Beant College of Engineering and Technology, Gurdaspur
rajeevbedi@rediffmail.com

[2]Department of Information Technology, Beant College of Engineering and Technology, Gurdaspur
infomaticmohit@gmail.com

[3]Department of Information Technology, Global Institute of Engineering and Technology, Amritsar
Tajindersngh635@gmail.com

[4]Department of Computer Science and Engineering, SSCET, Badhani
punjab.sabbi@gmail.com

[5]Department of Information Technology, Beant College of Engineering and Technology, Gurdaspur
amritpal_bcet@yahoo.co.in

## ABSTRACT

*Privacy Security of data in Cloud Storage is one of the main issues. Many Frameworks and Technologies are used to preserve data security in cloud storage. [1] Proposes a framework which includes the design of data organization structure, the generation and management of keys, the treatment of change of user's access right and dynamic operations of data, and the interaction between participants. It also design an interactive protocol and an extirpation-based key derivation algorithm, which are combined with lazy revocation, it uses multi-tree structure and symmetric encryption to form a privacy-preserving, efficient framework for cloud storage. [2] Proposes a framework which design a privacy-preserving cloud storage framework in which he designed an interaction protocol among participants, use key derivation algorithm to generate and manage keys, use both symmetric and asymmetric encryption to hide the sensitive data of users, and apply Bloom filter for cipher text retrieval. A system based on this framework is realized. This paper analyzes both the frameworks in terms of the feasibility of the frameworks, running overhead of the system and the privacy security of the frameworks.*

## 1. INTRODUCTION

Cloud storage provides scalable and Quality of service guaranteed resources for storage, users can store and compute their data from any location at anytime by a device which can be connected with Internet to visit that cloud. Besides these powerful advantages of cloud Storage, however, many people and companies is still feel hesitant to store their data in cloud. The reason behind this hesitancy is the fear of people and companies regarding loss of control on their data because there are some incidents of data loss and data leakage which make people to think about it. E.g. a cloud storage-provider named Linkup lost his business last year after losing 45% of stored client data due to an error of a system administrator [3]. In 2007, criminal's targeted Salesforce.com cloud service provider, and steal customer emails and addresses using a phishing attack [4]. Even Google's Docs was visited by unauthorized attacker, which caused data leakage [5]. Therefore, cloud storage providers must consider the privacy issue in priority. A lot of people doing work on outsourced storage. [6] Developed a Privacy-Preserving electronic health record system. On the basis of Symmetric and Asymmetric encryption, it developed two key derivation schemes and compared the advantages and disadvantages of these key derivations. But main drawback of this is it did not consider the effects of change of user access right and the run time operations of data time which greatly influence the effectiveness of key derivation according to the analysis of the following sections. [7] Developed a model named PDAS for preserving privacy and integrity of aggregate query results. It supports privacy protection by dividing the owner's database into M sections and sending a section to a service provider. Any N of them can cooperate to recover the entire database, but any smaller group cannot. Main drawback with PDAS is that it didn't encrypt the data, so in this case the service provider can get the full database by getting partial information. Another drawback is that it demands many service providers to cooperate, which is not realistic. [8]-[10] describes how to do some special calculations on encrypted databases, e.g. KNN(k-nearest neighbor), Boolean queries and keyword based queries etc. Main drawback with these is that they didn't provide a framework for data storage and access, and even didn't consider key management, dynamics of access right and data. [11] Focused on cloud data storage security by distributing the cloud data file F duplicable across a set of a=b+c distributed servers. A (b+c,c) Reed-Solomon erasure-correcting code was used to create c redundancy parity vectors from b data vectors in such a way that the original b data vectors can be reconstructed from any m out of the b+c data and parity vectors. By placing each of the b+c vectors on a different server, the original data file can survive the failure of any c of the b+c servers without any data loss. It will be the future works of my research. [12] Proposed a scheme for efficient and secure access of outsourced data. It makes data index by binary tree, generates and managed keys by key derivation, it also deals with the dynamics of access right of user and data by over-encryption and/or lazy revocation. Its main drawback is that binary tree structure couldn't reflect the logical relation fully regarding organization of owners data; but this will increase the communication overhead as changing the user's access right would make other user whose access right doesn't change to update certificate, and it also occupy more storage space to store a control block on service provider which is uneconomical. Even it didn't consider how the dynamics of access right and data influences the effectiveness of key derivation. It didn't cover how to avoid collusive attack in which the revoked users cooperate with a cloud storage service provider.

The above analysis shows a privacy-preserving, efficient cloud storage framework is needed urgently. In this paper I am presenting the complete analysis of two Privacy Preserving cloud storage frameworks developed by [1] and [2]. In framework I, service providers and data owners manage data and build data index by multi-tree; for generating and managing keys, extirpation-based key derivation algorithm is designed to solve the ineffectiveness of key derivation; it deals with the dynamics of access right and data by lazy revocation; it ensures data confidentiality by symmetric encryption.

In framework II cipher text based retrieval means service provider can be an agent of data owner to retrieve the owner's data according to the user's query, and it doesn't know the content of the data and the query because they are encrypted, which protects the privacy of owner and user well is used. In this framework the owner is relieved from the overhead of data management, which reflects the main advantage of cloud storage.[13] Proposed symmetric encryption-based cipher text retrieval technique which support the owner to retrieve his own data and not allowed others to retrieve his data. [14] proposed an Asymmetric encryption-based cipher text retrieval scheme. In this the owner encrypted some keywords about his data in this the service provider support the owner to retrieve his own data by the keywords and it didn't not allow others to retrieve the data. [15] Proposed an Asymmetric encryption-based cipher text retrieval scheme which is used to help user M to put user N's data in service provider and only N can retrieve it. [16] Proposed a scheme based on bloom filter to retrieve data which matched with a Boolean query. But it was not fit for cloud storage because it needs the data owner to deal with the query of users and it always has a possibility of a false positive which is fatal to that situation. All, the above references are special cases of cloud storage and can't satisfy the demands of data sharing of cloud storage. Through the above analysis, we can see that a privacy preserving cloud storage framework supporting cipher text retrieval is needed urgently. In next sections of this paper, First of all I presented Privacy Preserving cloud storage framework I and framework II then several key issues in framework I and framework II and then performance evaluation of framework I and framework II.

## 2. Privacy-preserving Cloud Storage Framework I

Figure 1 shows different functional modules of data owner, user and cloud storage service provider and interaction between them. The dashed lines described the functional correspondence of connected parts. The interaction protocol is as following:

1) Owner (A) sends data block b and which is encrypted by key $k_b$ to Cloud Service Provider(C) for storage. And EA indicates the encryption algorithm, $k_{ac}$ is the key between A and C, $t_{mod}$ reflects the time of last update of the data block, MAC (Message Authentication Code) is used to verify the integrity of message:

$MSG_{AC}$= {A, C, $EA_{ac}$ (A, C, $EA_{kb}$ ($data_b$), $t_{mod}$, MAC)}

2) User (U) requests data blocks from A. And $k_{ua}$ indicates the key between U and A, req_index is the index of request, and keyword reflects what the user is interested in:

$MSG_{UA}$= {U, A, $Ek_{ua}$ (U, A, req_index, keyword, MAC)}

3) A verifies U's identity firstly, and searches on index and finds the data blocks which have the keyword and are satisfied the access control matrix after the verification is passed. Then, A sends the minimum key group $key_{min}$ of those data blocks and the certificate (certificate). Certificate includes the minimum number group $datanumber_{min}$ of those data blocks, kac is the key between A and C, $t_{certificate}$ indicates when the certificate is generated, and AR records the update times of the user's access right:

Certificate= {$E_{kos}$ (U, req_index, datanumbermin, $t_{certificate}$, AR, MAC)}

$MSG_{au}$= {A, U, Ekuo (A, U, req_index, keymin, certificate, MAC)}

4) U sends the certificate to C and asks for returning of those data blocks:

   $MSG_{UC}$= {U, C, O, req_index, certificate}

5) C tests the certificate. If it is legitimate, C returns those requested data blocks:

$MSG_{cu}$={C, U, req_index, Eki ($data_{ki}$) ||…|| $E_{kt}$ ($data_{kt}$), MAC}

6) U gets the data blocks, computes the keys of the data blocks from $key_{min}$ by key derivation algorithm, and then decrypts the data blocks.

There are two points about the framework needed to explain: first, the granularity of data block can be changed according to data owner's requirement, for example, a file or a 128K size data block. For the simplicity, the granularity of data block is file in the paper. Second, the files will not be encrypted if they needn't be kept secret.

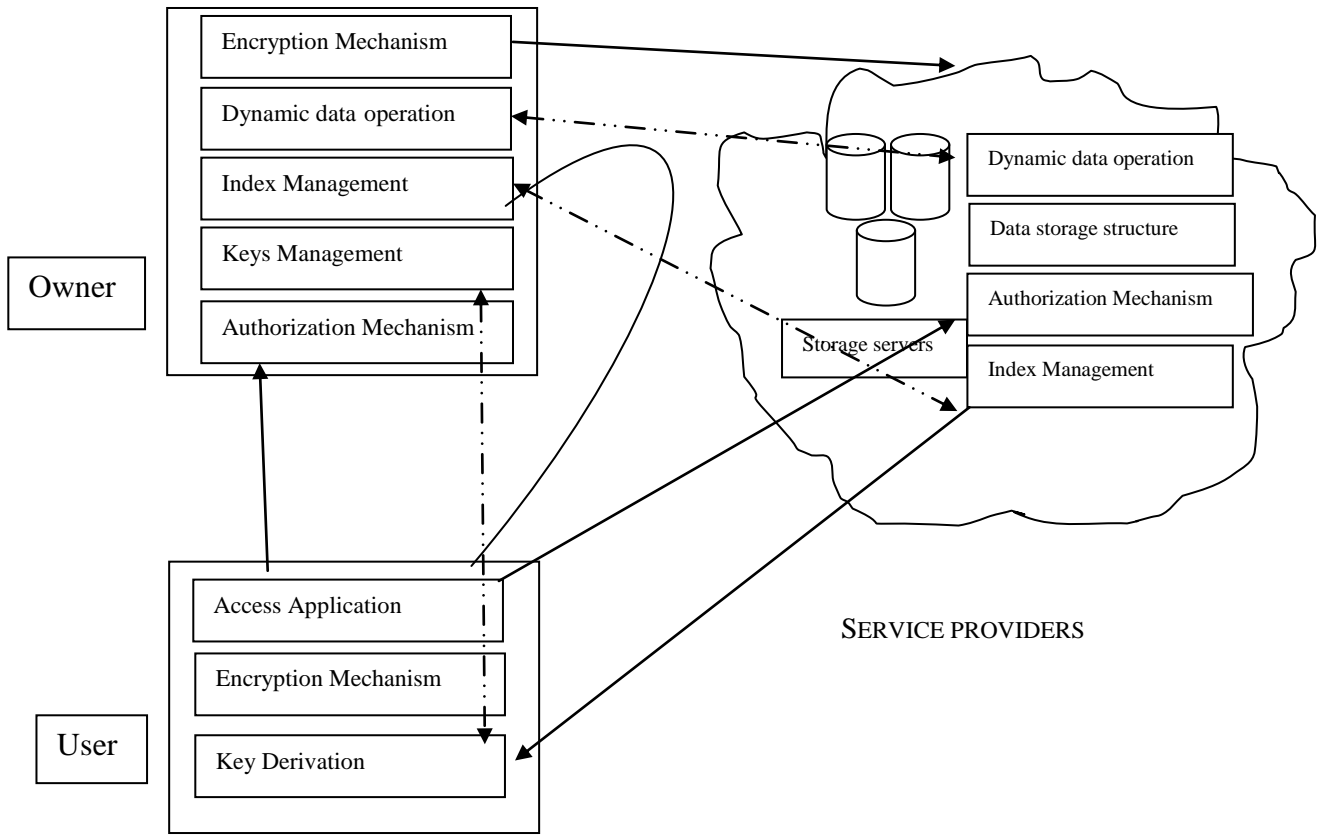

Figure 1 Privacy-preserving Cloud Storage framework I

## 3. Privacy Preserving Cloud Storage Framework II

Privacy-preserving cloud storage framework Supporting cipher text retrieval is showed in figure 2, which reflects the functional modules of data owners, users and cloud storage service providers and the interaction among them. The dashed line describes the functional correspondences of connected parts. The interaction protocol is as following:

1) Owner (A) chooses a root key $KEY_{root}$ for file encryption by symmetric encryption, a pair of keys ($k_{pub}$, $k_{pri}$) for keywords encryption of file by asymmetric encryption. Before $file_i$ is sent to Cloud Service Provider(C), owner generates the key $k_i$ of $file_i$ by key derivation algorithm and encrypts $file_i$. Then he encrypts keywords {$kw_1$, $kw2$,…,$kw_n$} by $k_{pub}$ and produce Bloom filter $BF_i$. At last, he sends encrypted files to service provider as following:

$MSG_{ac}$={A,C,$E_{kos}$(A,C,$E_{k1}$($file_1$)||$BF_1$||…||$E_{ki}$($file_i$)||$BF_i$,tmodified, MAC)}

And E indicates the symmetric encryption algorithm, $k_{ac}$ is the symmetric key between owner and service provider, $t_{mod}$ reflects the time of last update of the file, MAC (Message Authentication Code) is used to verify the integrity of message.

2) User (U) requests access authorization from owner. And kua indicates the symmetric key between user and owner, $request_{Id}$ is the serial number of request:

$MSG_{Ua}$= {U, A, $E_{kuo}$ (U, A, requestId, MAC)}

3) Owner verifies user's identity firstly, and searches on access control list to determine the files which can be accessed by user, then sends the minimum key group $key_{min}$ of those files and the certificate (cert) to user. Certificate includes the minimum number group $number_{min}$ of those files, $k_{os}$ is the symmetric key between owner and service provider, $t_{cert}$ indicates when the certificate is generated, and AR records the update times of the user's access right:

Cert= {$E_{kos}$ (U, $number_{min}$, $t_{cert}$, AR, MAC)}

$MSG_{au}$= {A, U, $E_{kuo}$ (A, U, requestId, $number_{min}$, $key_{min}$, cert, MAC)}

4) User sends the certificate to service provider and asks for some files which contain the keyword. AE indicates the

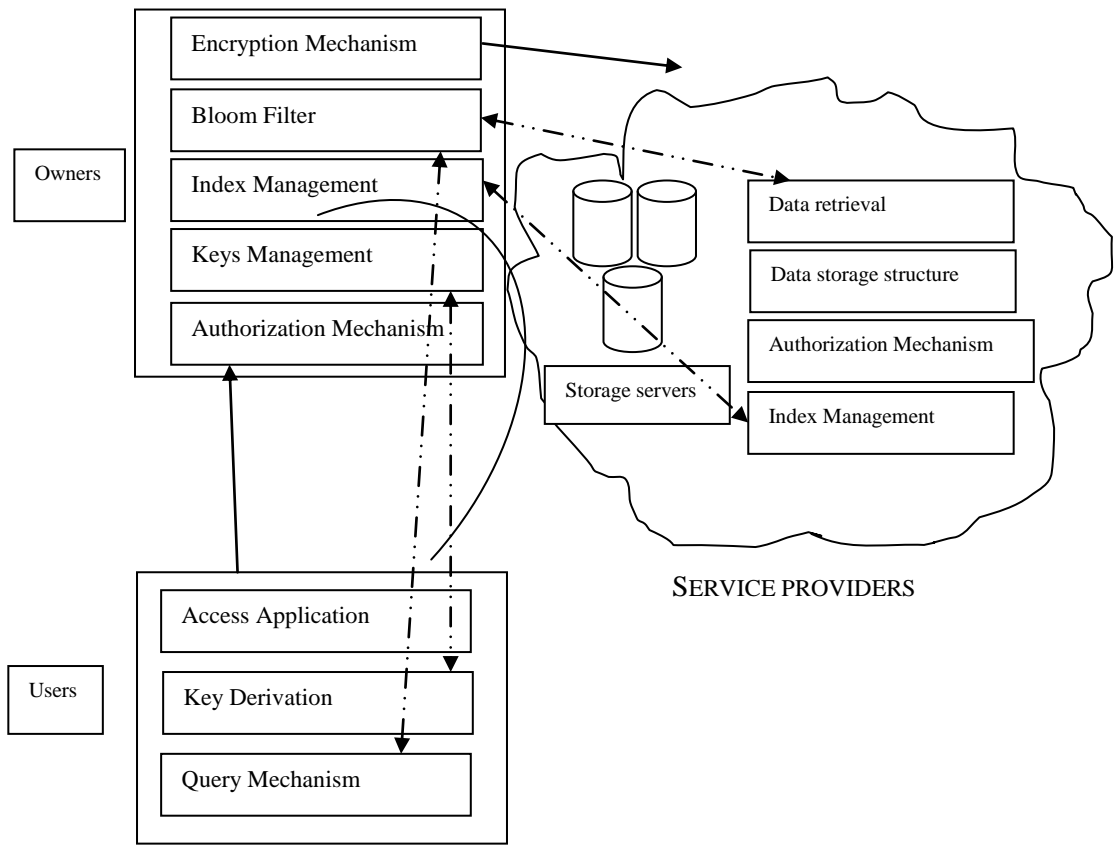

Figure 2 Privacy-preserving Cloud Storage frameworks II

Asymmetric encryption algorithm which is used to encrypt keywords by owner and kpub is the public key of owner:

$MSG_{UC}$= {U, C, A, $request_{Id}$, $AEk_{pub}$ (keyword), cert}

5) Service provider tests the certificate. If it is legitimate, service provider returns those requested files. $E_{ki}$ (file i) is the file which is encrypted by owner, and service provider never encrypts or decrypts owner's files.

$MSG_{cu}$={C, U, $request_{Id}$, $E_{ki}$ (file i) ||…|| $E_{kt}$ (file t), MAC}

User gets the files, computes the keys of the files from $key_{min}$ by key derivation algorithm, and then decrypts the files. Of course, the files will not be encrypted if they needn't be kept secret.

## 4. Several Key Issues in Framework I

### 4.1 Data Organization Structure

This framework uses multi tree based data organization structure for organizing files of Owner. This multi tree structure is automatically generated by client software when these files are going to be stored in the servers of cloud service provider. Figure 3 shows an example.

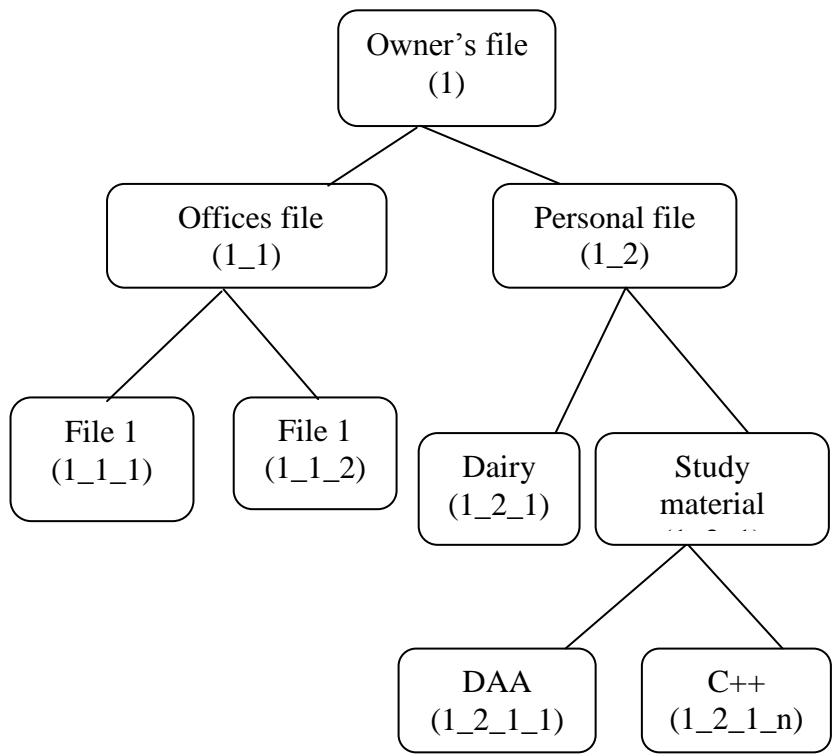

Figure 3 Multi-tree based Data Organization Structure of Owner's Files

In this files are basically the leaf nodes and non-leaf nodes represent folders or different categories of files. The contents and name of the file is encrypted by owner as file_number$$Ek_{file\_number}$ (file_name), for example 1_2_1$$Ek_{1\_2\_1}$(Diary), before he sends the file to cloud service provider. So it prevents cloud service provider knowing the content and name of the file, which provides the privacy of owner. When the cloud service provider receives the file, it will construct an index for every owner according the file's number, which basically speeds up search on data.

**4.2. Key Generation and key Management**

For better access to data, every file must have different key. So framework I uses symmetric encryption to reduce the burden of encryption and decryption. Key derivation [17] can be used for key management. In this Owner chooses a 128-bit key as root key by a random function, then produces sub key by the following formula: knum=hash(kpar||number||kpar), and hash( ) is a public hash function. Owner only needs to store the root key, which is not only convenient to key management, but also saves the owner's storage space. When a user asks for some files which are satisfied with the keyword, the owner will return the minimum key group from which all requested files' keys can be derived and other unauthorized files' keys can't. Key derivation can reduce the communication overhead of participants efficiently. But the effectiveness of key derivation will be harmed in some case: when the access right of a user is changed, owner must use a new key to encrypt the files if owner don't want the user to access the files which could be accessed by the user before. The new key can't compute by Knum=hash(kpar||number||kpar), and the framework generates the new key by choosing a 128-bit number randomly. When there are a lot of files using new key or every penultimate level directory has a file using new key, the effect of key derivation is the same as the situation where key derivation is not used, namely owner must return N keys if there are N requested files.

To solve this problem, we design extirpation-based key derivation algorithm: owner labels the node with "update" which will use a new key because of the change of user access right in the index tree, and creates a new node in update tree. The new node has the same number with the original node and has a new key. The course is shown in figure 3. When the node needs to update the key again, it can change the key of the node in update tree. When user requests some files, owner will compute the minimum key group by extirpation-based key derivation algorithm. The algorithm is as following:

public string ext_keyderivation(Node[ ] nodes，int t)

{

if (t = = 0){ *//when the key of a node is updated first time, and the node is represent as nodes[0]*

nodes [0].setUpdated( );

Node uNode=new Node(nodes[0].number,keyRandom()); *//uNode is updated node*

updateTree.addNode(uNode);

}

else if(t = = 1)

{       *//when the key of a node is updated non-first time, the node is represent as nodes[0]*

String key=Nodes[0].getUpdatedNode().createNewKey();

encrypt (file, key);

```
}
else{    // compute the minimun key group
String key_min=" ";
Node pNode=null;
for(int i=0;i<nodes.length;i++)
{
if(nodes[i].updated==1)
key_min=key_min+nodes[i].getUpdateNode().getKey();
else
{
parentNode=findParentNode(i,nodes);
if(parentNode!=null)
{
key_min=key_min+pNode.getKey();
Node[] newNodes=nextNodes(nodes,pNode);
String s=extirpated_keyderivation(newNodes, 3);
key_min=key_min+s;
}
Else
 key_min=key_min+nodes[i].getKey(); i++;
}
}
}
 return key_min;
}
```

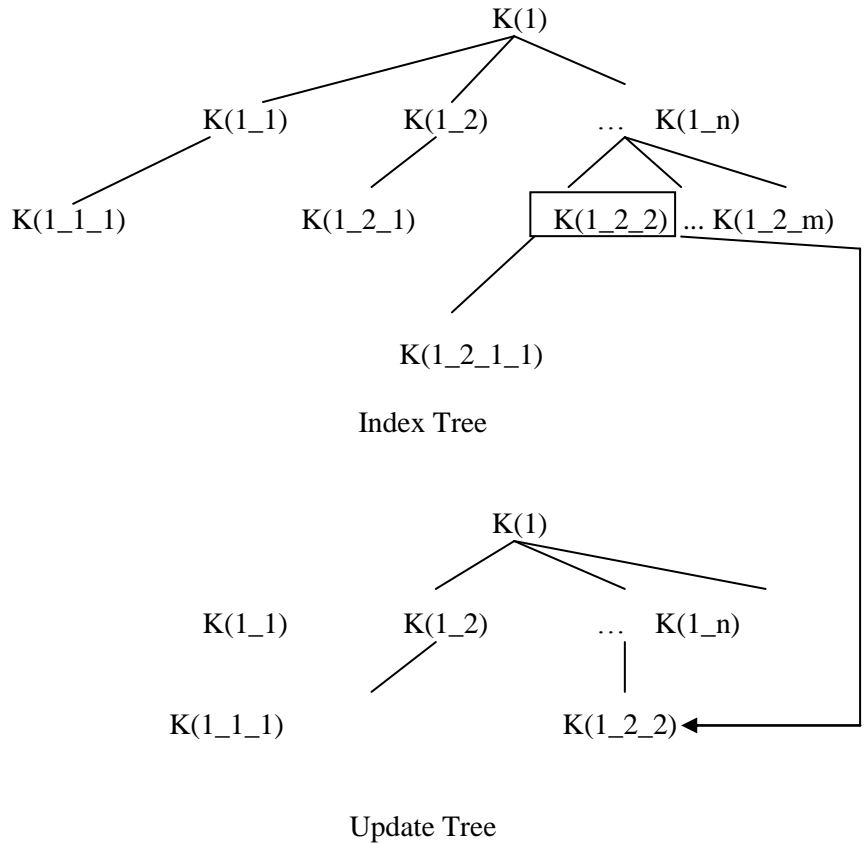

Figure 4 the Correspondence between Index Tree and Update Tree

There is an example. When owner updates the key of file 1_2_2, he will get the file from service provider and decrypt it by the old key firstly. He asks service provider to delete the file and mark the node 1_2_2 with "updated". Then he encrypts the file's content and name with new key of the node in update tree, and sends the encrypted file to the service provider. When an authorized user requests the files which are in the folder 1_2, the owner searches the index tree and returns the minimum key group which includes key1_2 and key1_2_2。From the effect of the algorithm, node 1_2_2 seems to be extirpated from the index tree. The algorithm can reduce the number of returned key effectively.

### 4.3 Access Right Change

First, service provider builds an access right updating linked list **updateAR[Owner_id]** for every owner, and the node in linked list has two properties: **node.id** is the number of user, and **node.times** indicates how many times the access right of the user was updated. After Owneri updates the access right of Userj，he sends the update massage to ServiceProviderk with the number of Userj. ServiceProviderk receives the massage and searchs the linked list updateAR[i]. If there is a node with node.id=j, then node.times++; otherwise ServiceProviderk inserts a new node into updateAR[i] and set node.id=j and node.times=1. When Userj requests files from ServiceProviderk, ServiceProviderk checks whether there is a node with node.id=j in updateAR[i]. If there is not such a node, ServiceProviderk returns the files; if there is such a node, ServiceProviderk will check whether node.time is equal to cert.AR. If node.time is equal to cert.AR, it will return the files; otherwise it will refuse to return the files and remind the Userj that his certification has expired. The above operations prevent revoked user getting files from service provider. Of course, a revoked user can steal files when they are transmitted. There are two methods to solve the problem: one is over-encryption[18] and the other is lazy revocation[19]. Over-encryption asks the service provider to encrypt the files before they are transmitted, which can prevent revoked user getting the files, but not all service providers are willing to provide such a service and encrypting a batch of files increases the economic burden of owner. Lazy revocation doesn't need owner and service provider to do anything before the file is updated because the stolen file is the same as the file which the revoked user had authorization to access. The framework adopts lazy revocation.

### 4.4 Dynamic Operations of Data

Owner has three dynamic operations on data: addition, deletion and update. When owner wants to add a new file, he will find a new number from the index tree according to the logical relation and compute the key by knumber=hash(kparent||number||kparent), and then encrypt the file and store it in service provider. When owner wants to delete a file, he will send a delete message to service provider to delete the file, then mark the node of the file in the index tree with "deleted". When there is a new file which wants to use the number of the deleted file, it will be treated as an updated file. When the file is updated, the key is valid if there is not a revoked user who could access the file before. Otherwise we need to do the following operations: owner marks the node of the file in index tree with "updated", and inserts a new node with same number and new key into update tree. Then he encrypts the content and name of the file with new key, and sends the encrypted file to service provider. Suppose tmodified indicates when the file was modified lasted and tcert indicates when the user's certificate was created. When an user requests the file, service provider compares tmodified and tcert, cert.AR and node.times of node whose node.id is equal to the user's number in updateAR[owner_id]. If tmodified>tcert and cert.AR==node.time, the user is an authorized user whose key is old, so service provider will return the file and remind him to get a new key; if tmodified≦tcert and cert.AR==node.time, the user is an authorized user who's key is new, so service provider will return the file; if cert.AR<node.time, the user is an revoked user, so service provider will refuse to return the file to him.

## 5. Several Key Issues in Framework II

### 5.1 Data Organization Structure, key derivation and management

Owner organizes his files in accordance with some logical relations. For reflecting the logical relations, the framework constructs the file index by multi-tree. Before those files are stored in service provider, the client software of owner will generate multi-tree index automatically according to their logical relation. Figure 2 shows an example. In such an index, only leaf nodes correspond to files, and non-leaf nodes represent folders or categories of files. Owner encrypts the content and name of a file and changes its' name as file_number$$_{Ekfile\_number}$(file_name), for example 1_2_1$Ek1_2_1(Diary), before he sends the file to service provider. The pretreatment prevents service provider from knowing the content and name of the file, which protects the owner's privacy. The service provider will construct an index for every owner according the files' numbers, which can accelerate search on data. To have a flexible and fine-grained access control, every file has a unique key. The framework uses symmetric encryption to reduce the burden of encryption and decryption. But how to manage numerous keys? Key derivation can be used to solve the problem. Owner chooses a random 256-bit key as root key, then produces sub key by the following formula: keynumber=hash (keyparent||number||keyparent) and hash () is a public hash function. When a user asks for the access authorization, the owner will return the minimum key group from which all authorized files' keys can be derived and other

unauthorized files' keys can't. For example, if user is authorized to access the files under the folder 1_2, owner just returns the key1_2. User can compute the keys of the files from key1_2 by key derivation algorithm. Key derivation not only facilitates key management, but also saves the owner's storage space and reduces the communication overhead of participants efficiently.

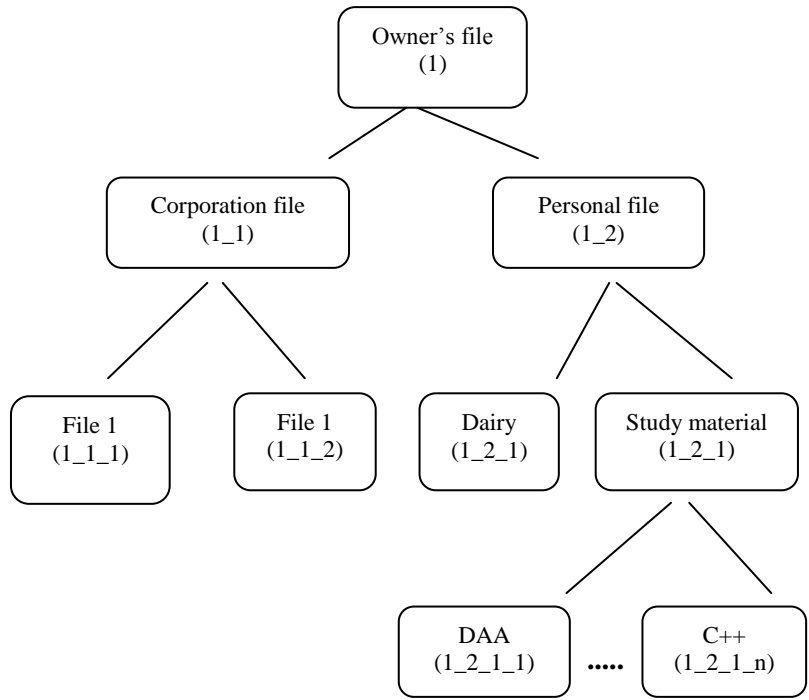

Figure 5 The Index Structure of Owner's Files

**5.2 Cipher text Retrieval Based on Bloom Filter**

In cloud environment, owner stores his files in storage servers of service provider. Anybody who gets the authorization from owner can retrieve files by the help of service provider. Service provider will retrieve the owner's files according to the owner's authority scope and the user's query. The advantage of the design is that service provider undertakes the job of file retrieval, which reduces the computing pressure of owner and is convenient for file sharing. When owner's files are stored in plaintext and the query is expressed in plaintext too, file retrieval is easy. But when file and query are encrypted, it is not easy to retrieve files. Our scheme can solve the problem well. There are three key steps in our scheme: keyword extraction, Bloom filter generation and keyword retrieval.

(1) Keyword Extraction  Owner finds some keywords

to describe a file. When there are a lot of files, it is usually a miscellaneous and toilsome job. So we design a client software to extract keywords from filename of a file according the language character. For example, there is a file named "The Storage of Cloud Computing", and the keywords will be "storage", "cloud", "computing".

(2) Bloom Filter Generation: Owner chooses a pair of

keys(Kpub,Kpri), and the parameters of the Bloom filter, such as the number of hash functions k and the size of a bit array m. Then he encrypts keyword$_{ij}$ of file$_i$ by kpub, namely KW$_{ij}$=AEkpub(keyword$_{ij}$), and AE() is asymmetric encryption. So File$_i$ has an encrypted keyword set:{KW$_{i1}$, KW$_{i2}$,…, KW$_{in}$}. Every element of keyword set is calculated as following:

y1=hash1(i||KW$_{ij}$),y2=hash2(i||KW$_{ij}$),… yk=hashk(i||KW$_{ij}$),

and then array bits of BF$_i$ at position y1, y2,…, yk are set to

1. Concatenation with the file number is necessary to make the bit pattern of Bloom filter BF$_i$ and BF$_m$ completely different even if the keywords of them are the same. At last, owner stores encrypted file$_i$ and BF$_i$ in service provider.

(3) Keyword Retrieval: User hopes to search the file

whose keyword is str. He requests access authorization from owner firstly. Owner checks the identity of user and determines his access scope. For example, owner authorizes user to access the files under the file folder 1_2 in figure 2, so he puts numbermin=1_2 into certificate and sends it back to user. User encrypts str by owner's public key, namely

w=AEkpub(str), and then sends w to service provider with the certificate. After receiving the query, service provider gets the numbermin from user's certificate, and then searches Bloom filter of every file in the scope of numbermin: array bits at positions h1(file_number||w),…, hk(file_number||w) are checked. If any selected bit is 0, str is definitely not a

keyword of the file. On the other hand, if all the checked bits are 1, then w is considered as a keyword of the file. Through the above steps, service provider can find the files whose keyword is str even it doesn't know the content of the file and query, thereby ciphertext retrieval is realized. Adopting Bloom filter, owner needn't store real keywords in cloud, and he just store a bit array which carries the keywords' information, so it is efficient, safe and economic, which will be verified in the next section.

## 6. Performance Evaluation of framework I

### 6.1 Effectiveness of Extirpation-based Key Derivation Algorithm

To reflect the real cloud storage environment, experiment simulates the interactions among multiple users, multiple owners and multiple service providers. User requests files randomly, and owner changes user's access right and updates files randomly, too. Owner stores thirty files in different organization structures in service provider's server. Suppose the size of minimum key group of extirpation-based key derivation is **size1**, and the size of minimum key group of common key derivation is **size2**. By computing size1/size2, the effectiveness of extirpation-based key derivation can be verified, which is showed as Figure 6.

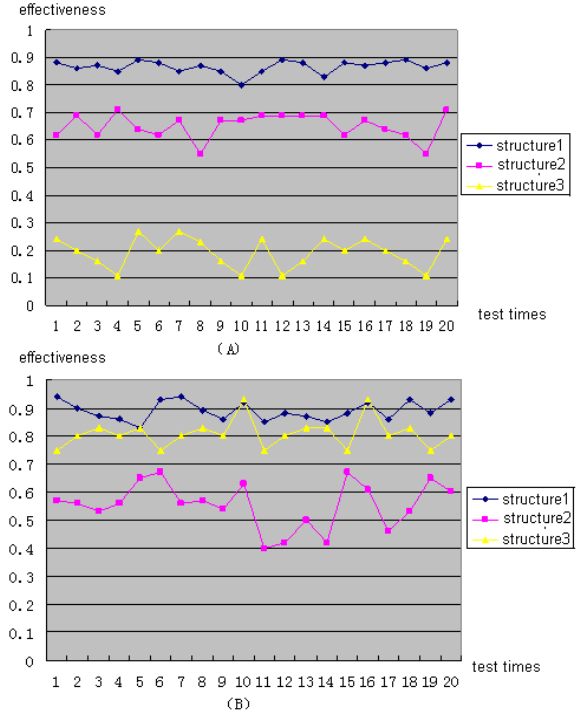

Figure 6 The Effectiveness of Extirpation-based Key Derivation Algorithm

The Effectiveness of Extirpation-based Key Derivation Algorithm Figure (A) shows the effectiveness when updating the same file in three different file organization structures; figure (B) shows the effectiveness when updating another file in the above three structures. From figure (A) and (B), we can draw conclusions as following: (i) extirpation-based key derivation algorithm is very effective because size1/size2<1; (ii) the organization structure of files has a direct influence on the effectiveness of the algorithm; (iii) the position of the updated file has a direct influence on the effectiveness of the algorithm; (iv) when a file organization structure is fixed, the effectiveness of the algorithm fluctuates surrounding a value. The reason is that the effectiveness of the algorithm is 2/n if there are n files in an folder which has a updated file. So the Effectiveness will fluctuate around 2/n.

### 6.2 Run-time Overhead of the system

Run-time overhead is measured from three aspects: communication, computation and storage overhead. The system in [12] is as a reference system. Suppose the amount of requested files is **nj** by **Userj**; **Owneri** has **mi** users, the length of user_id is **p**, the size of file t before and after it is updated is **lit** and **lit'**, the high of index tree is **h**, owner has **f** files; we adopt 128-bit key, **hash**() indicates the overhead of hash computing, and **E(t)** and **D(t)** is the computing overhead of encrypting and decrypting file t. The analysis is as following:As shown in Table 1, our system can reduce the communication, computing and storage overhead immensely.

Table 1 Performance comparison of two systems

| | | Our system Reference system | Our system Reference system |
|---|---|---|---|
| Communication Overhead | minimum key group | size1*128 | size2*128 |
| | minimum number group | size1*(h/2) | size2*(h/2) |
| | changing access right | p+1 | lit+lit'+(1/2) *mi *128 |
| | updating data | lit+lit'+(1/2)*mi*128 | lit+lit'+(1/2)*mi *128 |
| Computing overhead | key derivation (Owner) | nj *(1/2)*h*hash() | nj *(1/2)*h*hash() |
| | key derivation (Userj) | nj*(1/2)*h*hash() | nj (1/2)*h*hash() |
| | changing access right | ---- | E(t')+D(t) |
| Storage overhead | key | 128 | F*128 |
| | Control block | --- | 128+(h/2)+8 |
| | Update AR[i] | 1/2) *mi*8 | --- |

## 6.3 Privacy Security

From Figure 1, we can find there are several hidden dangers which could leak user's privacy: (i)during the course of files transmitting, outside attacker can steal the files by eavesdropping; (ii) inside attacker is easy to steal the files because the files are stored in service provider's servers; (iii)in collusive attacks, several revoked users or a revoked user and a service provider cooperate to steal the owner's files. Aimed at the first attack, attacker can't decrypt the file if he hasn't key. If he is a revoked user who has the key, he can decrypt the stolen file, but the file is the same as the file which he could has authorization to access before, which couldn't leak privacy. If the file was updated, the key of the file was changed too. So the key of the revoked user is invalid. To the second attack, owner encrypts the content and name of the files, and the keys are transmitted from owner to user. So the insider attacker has not the keys and couldn't decrypt the files. Against the third attack, there are two conditions: firstly, several revoked users cooperate to derive the key which couldn't be derived by the keys of one revoked user. Following the proof in [20], we can show that the attacks have to have a non-negligible advantage in breaking the hash function to accomplish this task. Therefore, the proposed approach is robust against such a collusive attack if the hash function is considered safe. Secondly, a revoked user and a service provider cooperate to steal the files. It only works when the revoked user has a key of a folder and owner adds a new file into the folder with the key computed by knumber=hash(kparent||number||kparent). At this time, service provider transmits the file to revoked user, user can derive the file's key by his key and decrypt the file. To solve the problem, owner adds a new file by the way of updating a file. Through the above analysis, the framework has a excellent privacy security.

## 6.4 Performance Evaluation of framework II

Our research group is designing and developing a campus-level cloud computing platform named "Qing Cloud". The project is composed of cloud computing and cloud storage. Based on the above framework, we developed a system of cloud storage by Java. Now we will judge the feasibility of the framework by analyzing the performance of Bloom Filter, the run-time overhead of the system and privacy security.

### 6.4.1 Performance of Bloom Filter

#### 6.4.1.1 False Positive Rate of Bloom Filter

Since the Bloom Filter is a probabilistic data structure, it always has a possibility of a false positive. The false positive rates are shown to be tunable by careful selection of parameters. There are three key parameters which can affect the false positive rate: the number of hash functions k, the size of a bit array m and the number of keywords n. We will use the following formula to compute the false positive rate c:

$$c = \left(1 - \left(1 - \frac{1}{m}\right)^{kn}\right)^k \approx \left(1 - e^{kn/m}\right)^k \quad (1)$$

Formula (1) is minimized for k=(m/n)*ln2, in which case it becomes:

$$c = \left(\frac{1}{2}\right)^k = (0.6185)^{m/n} \quad (2)$$

Suppose the false positive rate is less than 0.01%, then r is set to more than 14 and m should be more than 2*n.

#### 6.4.1.2 Overhead of Bloom Filter

We will analyze the performance of Bloom filter from computation, communication and storage overheads. In the framework, elliptic curve encryption algorithm(ECC) is used as the asymmetric encryption method which adopts 160-bit key. It encrypts keywords of files, and then transforms the cipher texts of a file's keywords into a Bloom filter. Suppose the false positive rate is less than 0.01%, there are five groups of keywords, which have different number of keywords and the keywords is generated randomly. The experiment is done in a computer with 1.86GH dual-core CPU and 2GB memory, and the result is

Table 2 Computation and Storage Overhead of Bloom Filter

| Type / Value Group | Storage and Communication Overhead | | | Computation Overhead | |
|---|---|---|---|---|---|
| | ECC(bit) | BF(bit) | BF/ECC | Owner(ms) | Service provider(ms) |
| 100 | 23855 | 2000 | 8.38% | 10.34 | 0.10 |
| 200 | 46615 | 4000 | 8.58% | 19.75 | 0.10 |
| 300 | 70768 | 6000 | 8.48% | 29.29 | 0.10 |
| 400 | 94382 | 8000 | 8.48% | 39.07 | 0.09 |
| 500 | 118143 | 10000 | 8.46% | 48.62 | 0.10 |
| 1000 | 234404 | 20000 | 8.53% | 97.13 | 0.10 |

Showed in table 1. From table 1, we can find that Bloom filter can reduce the communication and storage overheads. And at the same time, the computation overhead of Bloom filter is so small that it doesn't affect the performance of the framework, which is mainly used to calculate hash functions.

**6.4. 2 Run-time Overhead of the system**

Run-time overhead is measured from three aspects: communication, computation and storage overhead, as showed in Table 2. Suppose the amount of files which is authorized to access by Userj is nj, the amount of files which satisfies Userj's query is sj; Owneri has mi users, the size of encrypted filek is fk and the length of its Bloom filter is bfk, r is the amount of hash function in Bloom Filter, the high of index tree is h and the nodes in index tree occupies q bits, Owneri has p files, and the average amount of keywords of every file is g; we adopt 128-bit key, hash( ) indicates the computation overhead of hash function, E( ) and D( ) is the computation overhead of encrypting and decrypting file by symmetric encryption, AE( ) is the computation overhead of encrypting keyword by asymmetric encryption; len is the key amount in minimum key group generated by key derivation. The analysis is as following:

In the computation overhead and storage overhead of Table 1, (O) indicates that owner undertakes the overhead, the rest may be deduced by analogy. Our system reduces run-time overhead immensely by the following measures: 1) Storage overhead of owner, communication overhead of number group and key group is reduced greatly by key derivation; 2) According to the framework, file retrieval is done by service provider instead of owner, which relieves the computation overhead of owner; 3) Bloom filter can store multiple keywords' information in a bit, which saves the storage space and reduces communication overhead; 4) To use multi-tree structure, the length of file's serial number is shorter than or equal to the height of the tree, which reduces the communication overhead of number group.

**6.4.3 Privacy Security**

From figure 1, we can find there are several potential threats to users' privacy: (i)during the course of files transmitting, outside attacker can steal the files by eavesdropping; (ii) inside attacker is easy to steal the files because the files are stored in service provider's servers; (iii) several malicious users or a malicious user and a service provider cooperate to steal the owner's files, which is called as collusive attack; (iv)when the user queries, service provider may take a peep at the content of query which is the privacy of user. Aimed at the first attack, attacker can't decrypt the file if he hasn't key. If he is a revoked user who has the key, he can decrypt the stolen file, but the file is the same as the file which he had authorization to access before, which couldn't leak privacy. If the file is updated, the key of the file will be changed too. So the key of the revoked user is invalid. To the second attack, owner encrypts the content and name of the files by symmetric encryption, encrypts keywords of files by asymmetric encryption, and transforms encrypted keywords into a Bloom filter by hash functions. So the encrypted files and the Bloom filters are stored in service provider. The symmetric keys are transmitted from owner to user, the private key of asymmetric key is only known by owner, and the cipher texts of the keywords are not stored in servers of service provider. So the insider attacker couldn't decrypt the files and keywords. Against the third attack, there are two conditions: firstly, several malicious users cooperate to derive the key which couldn't be derived by the keys of one of them. Because hash function is a one-way function, the proposed approach is robust against such a collusive attack. Secondly, owner's certificate limits the scope of files which can be accessed by a user. And certificate is encrypted by the symmetric key Kos, which just can be decrypted by owner and service provider, so it can prevent the user from retrieving other files which is out of the scope. Because of the false positive rate of Bloom filter, service provider may return the files which don't meet the query. But the file is in the scope of authorization, so it won't leak owner' privacy. When service provider is in collusion with malicious users and retrieves files which is out of the authorized scopes, service provider can find the files meeting the query, but he can't decrypt those files because he haven't keys. If service provider wants to know the content of user's query, it can only do that by exhaust algorithm. Support there are eighty-five letters of which a filename can be made in alphabet, when there is a five-letter keyword, it spends 30ms to encrypt a string and retrieve Bloom filter one time by a computer with 1.86GH dual-core CPU and 2GB memory. So, 2.11 years will be spent to find out the five letter keyword, which is considered as difficult calculation. So the privacy of users can be protected. From the above analysis, the framework does well in privacy security.

Table 3 Run-time Overhead of the System

| Type | | Overhead |
|---|---|---|
| Communication Overhead | minimum key group | len*128 |
| | minimum number group | len*(h/2) |
| | file and bloom filter | $\sum_{k}^{p} 1(fk + bfk)$ |
| Computation Overhead | ciphertext retrieval(S) | (1/2)*nj*r*hash( ) |
| | key derivation (O) | nj*(1/2)*(1+h)*hash( ) |
| | key derivation (U) | sj*(1/2)*(1+h)*hash( ) |
| | file and keywords encryption(O) | p*(E( )+g*(AE( )+ r*hash( ))) |
| Storage Overhead | key(O) | 128 |
| | IndexTree(O) | p*h/2*q |
| | IndexTree(S) | p*h/2*q*mi |
| | file and bloom filter | $\sum_{k}^{p} 1(fk + bfk)$ |

## 7. Conclusion

In the conclusion, analysis of two frameworks named Privacy Preserving cloud storage framework I and Privacy preserving cloud storage framework II supporting cipher text retrieval. These frameworks constructs data index by multi-tree, generate and manage keys by key derivation, realize cipher text retrieval by Bloom filter. These frameworks support the interactions among multiple users, multiple owners and multiple cloud service providers, but only supports owner-write-user-read. This paper analyzes the feasibility of the frameworks from the performance of Bloom filter, runtime overhead of the system and privacy security. And the result verifies that the framework II is good at managing keys, protecting owner privacy and reducing communication, storage and computation overhead.

**Author**

I am working as a Assistant Professor in Computer Science and Engineering department at Beant College of Engineering and Technology, Gurdaspur, Punjab since 2004. I did my B.Tech. and M.Tech. from Punjab Technical University, Jalandhar. I am doing PhD. in Cloud Computing.

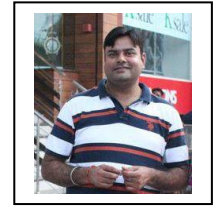

**Author**

I am working as a Assistant Professor in Computer Science and Engineering department at Beant College of Engineering and Technology, Gurdaspur, Punjab since 2004. I did my B.Tech. and M.Tech. from Punjab Technical University, Jalandhar. I am doing PhD. in Cloud Computing.

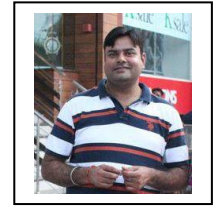